# New exact solution of the Navier–Stokes equations for turbulence in a compressible medium

Sergey G. Chefranov[1),2)] and Artem S. Chefranov[1),3)]

[1)]Obukhov Institute of Atmospheric Physics, Russian Academy of Sciences, Moscow, Russia

[2)]Physics Department, Technion-Israel Institute of Technology, Haifa 32000, Israel

[1)]schefranov@mail.ru, [2)]csergei@technion.ac.il, [3)]a.chef@bk.ru

## Abstract

A new exact solution of the Navier–Stokes equations is derived for the compressible medium flows which are far from equilibrium in the limit of extremely low shear viscosity and relatively large volume viscosity. This solution corresponds to the exact vortex solution in Euler variables for the three-dimensional Hopf equation and loses smoothness in a finite time. It is shown that the smoothness of the solution is conserved indefinitely in time if we take into account the contributions of small shear viscosity; it is also conserved in the case of homogeneous friction above a given threshold. The closed description of the evolution of statistical moments of velocity is obtained, thereby bypassing the closure problem in the theory of turbulence.

# 1. Introduction

For almost two centuries, possibly the last unsolved problem of classical physics associated with the puzzling problem of turbulence has continued to exist, a problem which has fundamental and practical importance not only for hydrodynamics, magnetic hydrodynamics and even hemodynamics, but also for many other branches of modern physics [1-3]. The absence of non-stationary solutions of the non-linear Navier–Stokes equations has led to the development of a statistical approach to the analysis of its solutions based on an infinite open-loop system of equations for the moments of the velocity field [1, 2]. Approximate approaches to solving the corresponding problem of closure of this system of equations were proposed in the works of Reynolds, Kolmogorov, Heisenberg and other prominent mathematicians and physicists.

Even the existence of such solutions over an unbounded time interval is an unsolved mathematical problem in the three-dimensional case, a problem which is of great practical importance in connection with the problem of predictability and stability arising in the numerical solution of the Navier–Stokes equations. These solutions are used in aerodynamic and hydrodynamic calculations for constructions of watercraft, aircraft and space technology, as well as for the prediction of ordinary and extreme weather phenomena. Therefore, in the year 2000 the Clay Mathematics Institute named the problem of the existence and uniqueness of solutions of the three-dimensional Navier–Stokes equations as one of the seven Millennium Prize Problems (www.claymath.org [4]). In [4], however, the problem is limited only to an approximation of an incompressible medium with a velocity field with zero divergence, that is $div\vec{u} = 0$.

In this paper, new exact non-stationary solutions to the three-dimensional Navier–Stokes equations for a compressible medium with $div\vec{u} \neq 0$ are obtained; these give the solutions for a generalization of the problem stated in [4]. This corresponds to the limit of extremely small shear viscosity as $\eta \to 0$ and to large volume viscosity with coefficient $\varsigma \gg \eta$. In this case, we use, instead of the equation of state, the representation for pressure of systems far from equilibrium in the form given in [2, 3], as justified by the condition (see (81.4) in [2] and (1.95) in [3]):

$$p = \varsigma \, div \vec{u} \qquad (1.1)$$

For the first time, based on a new solution to certain of the Navier-Stokes equations, an example is given which represents the possibility of a closed statistical description for a steady-state energy spectrum that determines the energy transfer between motions at different scales in a turbulent flow of a compressible medium.

We also show the possibility to avoid the collapse over a finite time of the solution via two mechanisms: (a) regarding non-zero shear viscosity $\eta \neq 0$, and (b) by taking into account the magnitude of the homogeneous friction with the coefficient $\mu$ which satisfies the relation:

$$\mu > \mu_{th} = 1/t_0 \qquad (1.2)$$

In (1.2), $t_0$ represents the finite interval of time for the collapse of the solution for $\mu = 0, \eta = 0$. In all numerical stimulations of the Navier–Stokes equations, cut-off procedures at spatial scales below $l < l_{min}$, or above $k_{max} \approx 1/l_{min} > k$ are used. This is one of the reasons for which problems with predictability problem arise. Indeed, this procedure corresponds to the introduction of an effective homogeneous friction with a coefficient $\mu = k_{max}^2 \nu$, where $\nu = \eta / \rho_0$ is the constant coefficient of kinematic viscosity. The choices of the cut-off scale $l_{min}$ are now found without any correlation with the initial conditions which determined the magnitude of $t_0$ in the right side of condition (1.2). Maintaining smoothness of a solution in the form of (1.2) allows us to establish a new basis for the elimination of the problem of predictability.

## 2. Exact solution of the Navier–Stokes equations

1. The Navier–Stokes equations for a compressible medium with constant shear coefficient $\eta = cons$ and volume (or second) viscosity coefficient $\varsigma = const$ have the form [1]:

$$\frac{\partial u_i}{\partial t} + u_j \frac{\partial u_i}{\partial x_j} = \frac{\eta}{\rho} \Delta u_i - \frac{1}{\rho} \frac{\partial}{\partial x_i}(p - (\varsigma + \frac{\eta}{3}) div \vec{u}) \qquad (2.1)$$

In (2.1), $p, \rho$ are pressure and density fields, respectively; $u_i; i = 1,..,n$ corresponds to the velocity field, and the indices range from 1 to $n$. In the following, cases of one-dimensional (1 D) $n = 1$, two-dimensional (2 D) $n = 2$ and three-dimensional (3 D) $n = 3$ spaces in (2.1) are considered.

Equation (2.1) must be considered together with the continuity equation:

$$\frac{\partial \rho}{\partial t} + div(\rho \vec{u}) = 0 \qquad (2.2)$$

The system of equations (2.1) and (2.2) is not closed, because, for example, in the 3D case, it contains four equations and five unknown functions, that is, three components of the velocity field, as well as the pressure and density fields. Usually, when considering an ideal compressible medium, the equation of state, which relates density and pressure, is used to close the system of equations (2.1) and (2.2). In this case, it is assumed that the time for establishing local thermodynamic equilibrium is much smaller than the characteristic time scale of the hydrodynamic process under consideration. However, in cases of turbulent flows with large Reynolds and Mach numbers, this assumption is not always fulfilled. In these cases the pressure field will depend on the velocity field and its derivatives. Such a dependence of pressure on velocity occurs even at the limit of small Mach numbers of approximations used for an incompressible medium having a constant density, when, for the closure of system (2.1) and (2.2), the condition of zero velocity field divergence is used:

$$div \vec{u} = 0 \qquad (2.3)$$

In the limit of large Mach numbers, the closure of (2.1) and (2.2) by use of the equation of state is all the more unjustified because that equation characterizes the state of thermodynamic equilibrium. Indeed, in [2], instead of the equation of state, equation (1.1), which relates the pressure to the velocity field divergence, is used (see (81.4) in [2] and (1.95) in [3]). Furthermore, using equation (1.1) instead of equation (2.3) for the case of a compressible medium leads to the closure of the system of equations (2.1) and (2.2).

2. In the limit $\eta \to 0$ and under the condition (1.1), the right-hand side of Equation (2.1) vanishes. As a result, the Navier–Stokes equations (2.1) are reduced to a generalization of the well-known one-dimensional Hopf equation for 2D and 3D cases [5]. Such a generalization of the Hopf equation describes the motion of liquid particles by inertia and has the form [5-7]:

$$\frac{\partial u_i}{\partial t} + u_k \frac{\partial u_i}{\partial x_k} = 0, i = 1,...,n \qquad (2.4)$$

As a result, the equation for the velocity field becomes independent of the continuity equation (2.2).

The solution of the Hopf equations (2.4) in Euler variables will simultaneously determine the solution of the well-known Helmholtz equation for the vortex field $\vec{\omega} = rot\vec{u}$. Indeed, after applying the operator for the curl to equation (2.4), the vortex Helmholtz equation for the case of a compressible medium and a nonzero divergence of the velocity field has the form:

$$\frac{\partial \omega_i}{\partial t} + u_k \frac{\partial \omega_i}{\partial x_k} = \omega_k \frac{\partial u_i}{\partial x_k} - \omega_i \frac{\partial u_k}{\partial x_k} \qquad (2.5)$$

For the Hopf equations (2.4), however, only solutions in the Lagrange variables [5–7] have long been well known.

In the case of a solution in Euler variables for the 1D case, Equation (2.4) is solved in [8, 9]. For solving the Helmholtz equations (2.5), only 2D and 3D cases are of interest. The solution of (2.4) and (2.5) for 2D and 3D cases was first obtained in [10] (see also [11 - 14], where the details of the derivation of this solution are given); it has the form

$$u_i(\vec{x},t) = \int d^n \xi u_{0i}(\vec{\xi}) \delta(\vec{\xi} - \vec{x} + t\vec{u}_0(\vec{\xi})) \det \hat{A} \quad (2.6)$$

In (2.6), $\det \hat{A}$ is the determinant of the matrices $\hat{A} \equiv A_{nm} = \delta_{nm} + t\frac{\partial u_{0n}}{\partial \xi_m}$, $\delta$ (delta) is the Dirac function and $u_{0i}(\vec{x})$ is the arbitrary smooth initial velocity field which corresponds to the finite integral of the energy $E_0 = \frac{1}{2}\int d^n x u_0^2(\vec{x}) < \infty$. The solution (2.6) preserve smoothness only over the interval of time $0 \leq t < t_0$, as fixed from the condition of the positive determinant $\det \hat{A} > 0$ where $t = t_0$ is the minimum time over which the determinant becomes zero, that is $\det \hat{A} = 0$.

For example, for the 1D case, $\det \hat{A} = 1 + t\frac{du_{01}}{d\xi_1}$ when $t_0 = \frac{1}{\max|du_0/dx|} > 0$.

In particular, with the initial velocity field is of the form $u_0(x) = a\exp(-x^2/2L_0^2), a > 0$, we get $t_0 = \frac{L_0\sqrt{e}}{a}$ when $x = x_{\max} = L_0$. For the 2D and 3D cases, the explicit definition of the minimum collapse time $t_0$ of the solution is associated with the solution of the quadratic and cubic equations, respectively. These equations follow from the condition that the determinant of a matrix $\hat{A}$ is zero, included in the definition of the solution (2.6) (see (3.7) in [11]).

The vortex field corresponding to the solution (2.6) and exactly satisfying the nonlinear Helmholtz equations (2.5) (see the proof of this in [11, 13]) has the form for the 3D case:

$$\omega_i(\vec{x},t) = \int d^3\xi (\omega_{0i}(\vec{\xi}) + t\omega_{0j}\frac{\partial u_{0i}(\vec{\xi})}{\partial \xi_j}) \delta(\vec{\xi} - \vec{x} + t\vec{u}_0(\vec{\xi})) \qquad (2.7)$$

In (2.7) $\vec{\omega}_0(\vec{x}) = rot\vec{u}_0$ is the initial distribution of the vortex field. In this case, the second term in parenthesis corresponds to the description of the well-known three-dimensional effect of stretching of the vortex lines [1].

For the density field satisfying equation (2.2), taking into account solution (2.6), we also obtain the exact solution in the form:

$$\rho(\vec{x},t) = \int d^n \xi \rho_0(\vec{\xi}) \delta(\vec{\xi} - \vec{x} + t\vec{u}_0(\vec{\xi})) \qquad (2.8)$$

In (2.8), $\rho_0(\vec{x})$ represents the initial distribution of the density field. Note that in the 2D case, the exact solution of the Helmholtz vortex equation (2.5) also has the expression (2.8) if we replace the distribution $\rho_0(\vec{x})$ in (2.8) by the initial distribution of vorticity $\omega_0(\vec{x})$.

While considering representations (1.1) for pressure we also obtain the exact solution following from (2.6) (see also [11, 13]):

$$p = \varsigma div\vec{u} = \varsigma \int d^n \xi \frac{\partial \det \hat{A}}{\partial t} \delta(\vec{\xi} - \vec{x} + t\vec{u}_0(\vec{\xi})) \qquad (2.9)$$

Thus, relations (2.6) to (2.9) give a new exact solution of the Navier–Stokes system of equations (2.1) and (2.2), based on the use of relation (1.1) for compressible medium flows with nonzero velocity field divergence. The resulting solution, however, is defined and retains smoothness only on a finite time interval $0 \le t < t_0$, which depends on the specific type of initial conditions (see (3.7) in [11]).

### 3. Regularization of the solution and the problem of predictability.

1. The exact solution of the Navier–Stokes equations (2.6) to (2.9) can be carried out, for example, when taking into account homogeneous friction. As is noted in the Introduction, homogeneous friction inevitably arises in any numerical calculations because of the need for a cut-off procedure which can significantly affect the predictability and stability [15].

To account for homogenious friction in the right side of equation (2.4), one must add a term $-\mu\vec{u}$ which corresponds to a friction force which is linear over velocity with a coefficient $\mu > 0$. This term is also obtained from the shear viscosity force which occurs in the Navier–Stokes equations, when it is assumed that $\nu\Delta\vec{u} \cong -\nu k_{max}^2 \vec{u} \equiv -\mu\vec{u}$, which corresponds to the procedure of cut-off at large wave numbers $k > k_{max} = 1/l_{min}$.

With such a modification of equation (2.4), it is also possible to obtain an exact solution of the Navier-Stokes equations, which follows from expressions (2.6) to (2.9) by replacing the time variable with a new variable

$$t \to \tau(t) = (1 - e^{-\mu t})/\mu \qquad (3.1)$$

The new term (3.1) is only variable over a finite interval $0 \leq \tau(t) \leq 1/\mu$. As a result, in the equations (2.6) to (2.9) the magnitude of the determinant $\det \hat{A}$ can never vanish if relation (1.2) holds. For example, for the above estimated value $t_0$ in the 1D case, we obtain the regularization condition of the corresponding solution in the form $\mu > \mu_{th} = \dfrac{a}{L_0\sqrt{e}}$.

2. Another regularization method providing smoothness of the solution that is conserved over time is the modeling of the shear viscosity action using a random velocity field $\vec{V}(t)$ having a given Gaussian probability measure. To do this, in equation (2.4) we replace $u_i \to u_i + V_i(t)$; this corresponds to the replacement $\vec{x} \to \vec{x} - \vec{B}(t), \vec{B}(t) = \int_0^t dt_1 \vec{V}(t_1)$ in the solution given by Equations (2.6) to (2.9). We take $\vec{V}(t)$ as a random Gaussian velocity field (such as white noise), with delta correlation in time; more precisely, with the correlations $\langle V_i(t) \rangle = 0; \langle V_i(t+\tau)V_j(t) \rangle = 2\nu\delta_{ij}\delta(\tau)$. In this case, the expression in the angular brackets denotes the statistical averaging over an ensemble of realizations of a random velocity field. As a result, the solution represented by equations (2.6), after averaging over the random velocity field, become regular, conserving smoothness over time, with the form:

$$\langle u_i \rangle = \int d^n \xi u_{0i}(\vec{\xi}) \left| \det \hat{A} \right| \frac{1}{(\sqrt{2\pi \langle \vec{B}^2(t) \rangle})^n} \exp\left[ -\frac{(\vec{x} - \vec{\xi} - t\vec{u}_0(\vec{\xi}))^2}{2\langle \vec{B}^2(t) \rangle} \right] \quad (3.2)$$

$$\langle \vec{B}^2(t) \rangle = 2nt\nu$$

The averaged representations, regular over any interval of time, for the vortex, density, and pressure fields (2.7) to (2.9) are obtained analogously.

If we consider the average of the Gaussian velocity field for the cases of finite correlations in time, expressions analogous to those in (3.2) are obtained, in which only needs to replace the expression for the dispersion $\langle \vec{B}^2(t) \rangle$, as a function of time in (3.2) by a new time dependence for this quantity, taking into account the specific form of correlation over time for a random field $\vec{V}(t)$.

Thus, the averaged representation of the solution (2.6) in the form (3.2) corresponds to taking into account the shear viscosity modeled using a random Gaussian velocity field introduced into the Hopf equations (2.4). When taking into account the homogeneous friction in (3.2), it is necessary to replace the time variable according to relation (3.1). In particular, in the limit of large time intervals $t\mu \gg 1$, this replacement corresponds to $t \to 1/\mu$ and the stationary limit of the solution (3.2), which is used in the next section to obtain a closed description of two-point moments and energy spectra.

## 4. Application. Turbulence energy spectrum

On the basis of solution (3.2) to the Navier–Stokes equations (2.1), it is possible to obtain a representation for all multipoint moments of velocity, vorticity, density and pressure fields, thus giving the solution of the main problem of turbulence theory [1] (see chapter 3.3, pages 176-177 in [1]). In [1] it is also noted that for the case of a compressible medium, "…this common problem is too difficult, and the approach to its full solution is not yet visible" [1]. Here we consider the example in the compressible case of the solution for two-point moments of velocity field and corresponding energy spectrum which are of the most importance in turbulence theory.

The correlation tensor for the n-dimensional (n=1, 2, 3) velocity field is defined as:

$$R_{ij}(\vec{l}) = \frac{1}{L^n} \int d^n x \tilde{u}_i(\vec{x}+\vec{l},t)\tilde{u}_j(\vec{x},t); \tilde{u}_i = \langle u_i(\vec{x},t) \rangle \qquad (4.1)$$

$$L^n = (\frac{1}{2\sqrt{\pi}})^n P_0^2 / E_0; P_0 = \int d^n x |\vec{u}_0(\vec{x})|; E_0 = \int d^n x \vec{u}_0^2(\vec{x})$$

Let us consider (4.1) for the solution of Navier–Stokes equations for the steady limit $t\mu \gg 1$, when in (3.2) we must make the replacement $t \to \frac{1-\exp(-t\mu)}{\mu}$ after the introduction of homogeneous friction with coefficient $\mu > 0$. The correlation tensor (4.1) in this case is reduced to the formula

$$R_{ij}(\vec{l}) = \frac{1}{(4\pi L^2 \langle \vec{B}^2 \rangle)^{n/2}} \int d^n \xi \int d^n \xi_1 u_{0i}(\vec{\xi})u_{0j}(\vec{\xi}_1) \det \hat{A}(\vec{\xi}) \det \hat{A}(\vec{\xi}_1) \exp(-\frac{(\vec{l}+\vec{a}(\vec{\xi},\vec{\xi}_1))^2}{4\langle \vec{B}^2 \rangle}) \quad (4.2)$$

In (4.2), vector $\vec{a}(\vec{\xi},\vec{\xi}_1) = -\vec{a}(\vec{\xi}_1,\vec{\xi}) = \vec{\xi}_1 - \vec{\xi} + \frac{1}{\mu}(\vec{u}_0(\vec{\xi}_1) - \vec{u}_0(\vec{\xi}))$; in accordance with (3.2) we also have $\langle \vec{B}^2 \rangle = 2n\nu/\mu$. From (4.2) the symmetry relation $R_{ij}^u(\vec{l}) = R_{ji}^u(-\vec{l})$ is obtained.

The spectrum energy tensor corresponding to (4.2) is defined as

$$F_{ij}(\vec{k}) = \frac{1}{(2\pi)^n} \int d^n l R_{ij}(\vec{l}) \exp(i\vec{k}\vec{l}) \qquad (4.3)$$

Indeed, the energy is given by:

$$e = \frac{1}{2}\int d^n k F_{ii}(\vec{k}) \equiv \int_0^\infty dk E(k) \qquad (4.4)$$

For example, in the isotropic 2D case there is the relation $F_{ii}(k)\pi k = E(k)$; for the 3D case $F_{ii}(k)4\pi k^2 = E(k)$ similarly arises.

From (4.3) and (4.2) it is possible to obtain a common representation:

$$F_{ij}(\vec{k}) = F_{ji}(-\vec{k}) = \frac{1}{L^n} I_i(\vec{k}) I_j(-\vec{k}) \exp(-k^2 \langle \vec{B}^2 \rangle);$$

$$I_i(\vec{k}) = \int d^n x\, u_{0i}(\vec{x}) \exp(i\vec{k}(\vec{x} + \frac{1}{\mu}\vec{u}_0(\vec{x}))) \det \hat{A}(\vec{x}) \qquad (4.5)$$

For example, let us consider (4.5) with n=2 for the 2D initial velocity field (in the $(r,\varphi)$ system of polar coordinates):

$$u_{0r} = 0; u_{0\varphi}(r) = a\exp(-\frac{r^2}{2L_0^2}) \qquad (4.6)$$

In the case n=2, we set $L = L_0$ in (4.5) and $\mu_{th} = a/L_0$ in (1.2) (by using Equation (3.7) from [11, 13] to obtain $\mu_{th} = 1/t_0$). Thus, from (4.5) and (4.6) it is possible to obtain:

$$F_{\varphi\varphi}(k) = \frac{I_\varphi(k) I_\varphi(-k)}{L_0^2} \exp(-4\nu k^2/\mu); \qquad (4.7)$$

$$I_\varphi(k) = a\int_0^{2\pi} d\varphi \int_0^\infty dr\, r \exp(-\frac{r^2}{2L_0^2})(1 - \frac{\mu_{th}^2}{\mu^2}\exp(-\frac{r^2}{L_0^2}))\exp(ikS(r,\varphi));$$

$$S(r,\varphi) = r\cos\varphi - \frac{\mu_{th}}{\mu} L_0 \exp(-\frac{r^2}{2L_0^2}) \qquad (4.8)$$

For the estimation of integral (4.8) in the limit of the large wave number $k$, we use the method of stationary phase when the conditions $\frac{\partial S}{\partial r} = \frac{\partial S}{\partial \varphi} = 0$ are satisfied:

$$r = r_0 = L_0\sqrt{2}\ln^{1/2}(\mu_{th}/\mu); \varphi = \varphi_0, ctg\,\varphi_0 = -\frac{\mu_{th} r_0}{\mu L_0}\exp(-\frac{r_0^2}{2L_0^2}) \qquad (4.9)$$

As an example, for the case $\mu < \mu_{th} = a/L_0$ and in the limit $1 - \frac{\mu_{th}}{\mu} << 1$, we obtain here the estimate of (4.8) for the inertial range of wave numbers (when $k^2 4\nu/\mu << 1$ in (4.7)):

$$1/L_0 << k << \sqrt{\text{Re}}/L_0; \text{Re} = L_0 U_0/\nu, U_0 = L_0\mu/4 \quad (4.10)$$

For the range of wave numbers (4.10) corresponding to (4.7), the two-dimensional (2D) energy spectrum is obtained:

$$E(k) = \frac{C_E}{k^3}; C_E = \frac{\pi^3 a^2}{256 L_0^2} \ln^{-3}(\mu_{th}/\mu) \quad (4.11)$$

This representation of energy spectrum (4.11) is obtained only on the basis of the exact solution (3.2) of the compressible Navier–Stokes equations (2.1) and has the same scaling exponent as usually considered in 2D turbulence theory [16] when also $E(k) \cong O(1/k^3)$ in the inertial range of wave numbers is obtained, but for the incompressible case and not directly on the basis of the solution of the Navier–Stokes equations.

## Conclusions

Thus, the new exact non-stationary solution of the Navier–Stokes equations (2.6) obtained in this work admits regularization in the form of (3.2), which ensures its smoothness over an unbounded time interval by taking into account small shear viscosity. As a result, a positive answer is obtained for generalizing the Millennium Prize Problem [4] for the case of the Navier–Stokes equations describing the flow of a compressible medium in unbounded space and having a finite energy integral, as required in [4].

The regularization condition is obtained also by considering homogeneous friction in the form of (1.2); this provides a basis for new methods for the resolution of the above-mentioned problem of predictability.

For the first time, a general representation for two-point moments of the velocity field and the corresponding expressions for the energy spectrum determining the processes of nonlinear strong interactions between motions of different scales were obtained directly from solving the Navier–Stokes equations. On the basis of the exact solution thus obtained, any single-point and multi-point moments of the velocity, vortex, density, and pressure fields can also be found. This gives a solution for the perplexing problem of turbulence in a compressible medium.

We are grateful to E. A. Novikov, E. A. Kusnetsov, P. Lushnikov, I. Procaccia, V. L'vov and G. Fal'kovich for discussions and attention to this work.

The study is supported by the Russian Science Foundation, Grant number: 14-17-00806P, and the Israel Science Foundation, Grant number: 492/18.